# The Z CamPaign: Year Five

*Mike Simonsen*
*AAVSO*
*49 Bay State Rd., Cambridge, MA 02138 USA*
*mikesimonsen@aavso.org*

**Abstract**

Entering into the fifth year of the Z CamPaign, the author has developed a website summarizing our findings which will also act as a living catalog of bona fide Z Cam stars, suspected Z Cams and Z Cam impostors. In this paper we summarize the findings of the first four years of research, introduce the website and its contents to the public and discuss the way forward into year five and beyond.

## 1. Introduction

Z Cam stars are active dwarf novae characterized by a feature in their light curves known as standstills, a period lasting days to months when the star shines at a brightness approximately 1.5 magnitudes fainter than maximum light and several magnitudes brighter than minimum. The author launched the Z CamPaign in September of 2009 with the primary intent to sort out which dwarf novae are in fact Z Cam stars and which are not. The additional science goals of the campaign are described in detail in Simonsen (2011).

## 2. Results

After four years of concentrated observation and data analysis the total number of bona fide Z Cam stars currently stands at 22 (Stubbings and Simonsen 2014). The number of impostors is 27 and the number of suspected Z Cams stands at 19, out of 68 possible Z Cam stars.

Through an extensive literature search we have determined the main cause of the confusion and misclassifications in the literature on Z Cam stars stems from the fact that the very definition of what a Z Cam is has changed from its origin in 1932 (De Roy 1932) to the present. The presence of standstills in the light curves of Z Cams as the defining characteristic of Z Cam-ness did not come into use until 1971 (Glasby 1971). Thus, many stars originally determined to be Z Cams no longer fit the modern day definition. The present day definition, which also stresses the presence of standstills in the light curves, is found in the *General Catalogue of Variable Stars* (GCVS)(Samus *et al* 2007-2009).

Observation and analysis have led us to determine the other normal characteristics of Z Cam stars. All the known Z Cams have orbital periods between 3.0 and 8.4 hours, outburst cycle times (the time between successive maxima) ranging from 10 to 30 days, and outburst amplitudes from 2.3 to 4.9 magnitudes in V. Some Z Cam stars have been shown to exhibit deep fading episodes, reminiscent of VY Scl type stars.

An unexpected outcome of the Z CamPaign was the discovery of a previously unknown, and still unexplained, behavior in the light curves of IW And and V513 Cas (Szkody *et al* 2013). This behavior was later also found in the newly verified Z Cam star, ST Cha (Simonsen *et al* 2014b).

One of the most significant results of the campaign is confirmation that at least nine of the bona fide Z Cams actually do go back into outburst from standstill. This poses challenges to theories attempting to explain the change in mass transfer rates often quoted as the underlying cause for standstills as well as models of accretion and outburst mechanisms in dwarf novae in general (Simonsen *et al* 2014a).

The reclassification of so many stars once purported to be prototypical Z Cam stars to other dwarf novae sub types has a significant impact on the veracity of research papers written in the past describing Z Cam characteristics based on samples of stars contaminated by non-Z Cam members (Meyer and Meyer-Hofmeister 1983; Szkody and Mattei 1984; Shafter, Cannizzo and Waagen 2005).

Todays star catalogs, such as The International Variable Star Index (VSX), (Watson *et al* 2006), GCVS (Samus *et al* 2007-2009) and Ritter and Kolb (2012) should also be brought into line with our current knowledge of these systems.

## 3. The Z Cam List

*The Z Cam List* is a website cataloging the current status of the known and suspected Z Cams as well as Z Cam impostors, stars listed in the literature as Z Cams or suspected Z Cams that are not Z Cam stars (Simonsen 2009).

Stars in each category are listed in order of right ascension in three separate tables and have link outs to individual pages featuring each object. The individual object pages contain information on the star's characteristics, light curves, an AAVSO finder chart and a complete list of references.

The site includes a home page, that provides an introduction to Z Cam dwarf novae and the Z CamPaign written by the author, a list of papers published as a result of the Z CamPaign, an article contributed by Fred Ringwald entitled *"Why Observe Z Cam Stars?"* and the story of the unexpected discovery, by Rod Stubbings, of OQ Carinae's Z Cam nature.

## 4. The Future of the Z CamPaign

The first point of emphasis for the Z CamPaign moving forward will be to determine the true nature of the remaining Z Cam suspects. Of the remaining suspects, two, PY Per and V426 Oph, hold the most promise for being classified as Z Cams. Of the remaining stars, only those with short SS Cygni type outburst cycles are likely to be Z Cams. The rest are probably UG, UGSS and NL variables.

*The Z Cam List* will continue to be updated indefinitely with our latest findings, and we will continue to publish the results of the Z CamPaign in peer-reviewed publications.

Continued long-term, nightly monitoring of the bona fide Z Cams is highly encouraged. We have only begun to examine the unpredictable and fascinating nature of these cataclysmic variables in greater detail. We still have more questions than answers regarding this class of variable stars. What causes standstills? What proportions of Z Cams go back into outburst from standstill, and how is that even possible? Do star spots play a role in bringing standstills to an end? Why are there so few Z Cams among the thousands of known dwarf novae? Do they represent a short period in the evolution of cataclysmic variables from one type to another? Is there a relationship or continuum between Z Cams, NLs and VY Scl type systems since they share some of the characteristics of these systems? Are some or all of the Z Cams 'hibernating novae' as put forth in Shara et al 2007 and 2012?

## 5. Conclusion

In the first four years of the Z CamPaign we have met or exceeded all the science goals laid out in Simonsen 2011. We have unambiguously classified of 49 out of 68 potential Z Cam systems, determining that of those only 22 are in fact Z Cams. We have improved the overall quality of the data available for these stars and filled the gaps in their light curves with high cadence, quality data, since late 2009. We have determined that at least 9 of the bona fide Z Cams do go into outburst from standstill, resolving a long-standing question. We made several 'serendipitous discoveries' in the course of the campaign, most notably the unusual behavior of IW And, V513 Cas, and ST Cha. We have published 8 papers in peer-reviewed journals and facilitated the publication of at least two more papers based on data acquired as a direct result of the Z CamPaign (Ringwald 2012, Ringwald and Velasco 2012).

The real value of this campaign, and others like it, lies in the fact that long-term monitoring of a select list of stars like this can only be done by amateur astronomers. Due to decreasing access to telescope time, professional astronomers generally prefer to focus on observing projects they can complete in a few nights or on much shorter time scales than several years or decades. Without the amateurs conducting these observations this data would not exist.

With the creation of *The Z Cam List* website there is now a place online where the most up to date information on Z Cam stars can be accessed from a single source.

## 6. Acknowledgements

We acknowledge with thanks the variable star observations from the *AAVSO International Database* contributed by observers worldwide and used in this research. This research has made use of NASA's Astrophysics Data System.